# Mössbauer Spectroscopy Determination of iron foreign phases in the Superconducting Systems; $RAsFeO_{1-x}$, $RAsFeO_{1-x}F_x$ and $Sr_{1-x}K_xFe_2As_2$

I. Nowik[*] and I. Felner

Racah Institute of Physics, The Hebrew University, Jerusalem, 91904, Israel

The recently discovered superconducting - spin density wave materials, containing Fe and As, have raised huge interest. However most materials prepared to date, suffer from a varying degree of content of foreign Fe-As phases, $Fe_2As$, $FeAs_2$ and $FeAs$, which can lead to wrong conclusions concerning the properties of these materials. We show here that Mössbauer Spectroscopy is able to determine quite easily the relative content of the foreign phases. This procedure is demonstrated by a study of seven samples of superconducting or spin density wave materials, prepared in three different laboratories.

## Introduction

Recently two new families of high $T_c$ superconducting materials have been discovered. One family has the composition $RAsFeO_{1-x}F_x$ where R stands for a rare earth and the fraction x can be composed also by vacancies. The other family is $Sr_{1-x}K_xFe_2As_2$ where Ba can substitute Sr. These systems display in some cases superconductivity (SC) at temperatures ($T_c$) extending above 50 K. All systems are derivatives of magnetic spin density wave (SDW) systems (for x=0). SC emerges when doping of a magnetic parent compound with holes (or electrons) is done, and thereby suppressing the magnetic order. Some researchers report that they observe also magnetic order, even below $T_c$. Macroscopic measurements (like susceptibility or resistivity), or microscopic measurements, which sense averages of various localities (like μSR), are not able to determine the presence of foreign phases. Even XRD measurements are not sensitive enough to disclose small amounts of foreign phases, in particular if they have XRD lines similar to those of the SC system. Since all these systems contain iron, and this is the only element, which can order magnetically at relatively high temperatures, then a small amount of an iron compound, which orders magnetically, can contribute to the illusion that the superconducting materials exhibit also magnetic order. Mössbauer spectroscopy of $^{57}Fe$ of these materials can discover the presence of such compounds. The compounds, which are most probable to be present, are $Fe_2As$ ($T_N$=353 K), $FeAs$ ($T_M$=77 K) and $FeAs_2$ ($T_M$< 5K) [1-3].

## Experimental details

Mössbauer spectroscopy studies were performed using a conventional constant acceleration drive and a 50 mCi $^{57}Co$:Rh source. The velocity calibration was done with a room temperature α-Fe absorber and the isomer shifts (I.S.) values are relative to that of iron. The observed spectra, measured at temperatures extending from 4.2 K to 300 K, were least square fitted by theoretical spectra including, in the necessary cases, the full diagonalization of the hyperfine interaction Spin Hamiltonian, starting with hyperfine interaction parameters, corresponding to those of the assumed foreign phases [1-3]. The compounds $CeAsFeO_{0.84}F_{0.16}$, $Sr_{0.9}K_{0.1}Fe_2As_2$ and $Sr_{0.6}K_{0.4}Fe_2As_2$ were prepared and kindly supplied by Texas Center for Superconductivity at the University of Houston, U.S.A. The compounds $SmAsFeO$, $SmAsFeO_{0.85}$ and $SmAsFeO_{0.9}F_{0.1}$ were prepared and kindly supplied by Prof. Zhi-An Ren, National Laboratory for Superconductivity, Chinese Academy of Sciences, Beijing, China and $LaAsFeO_{0.9}$ was prepared and kindly supplied by NPL, New-Delhi, India [4].

**Experimental results and discussion**

a)  The system $LaAsFeO_{0.9}$

$LaAsFeO_{0.9}$ was prepared in evacuated quartz tubes in a single step procedure [5]. This system exhibits SDW magnetic order at 95 K, however at 200 K it displays well-defined absorption lines, Fig. 1, corresponding to the pure $LaAsFeO_{0.9}$ system (the major singlet) and an additional quadrupole doublet, which was erroneously claimed to correspond to iron nuclei which have a vacancy in their immediate environment [4]. However the hyperfine interaction parameters (I.S.= 0.35 mm/s and quadrupole splitting QS=eqQ/2=1.88 mm/s) are identical to those of $FeAs_2$ [1]. The relative abundance of this foreign phase is 11±1 %. The effect of a local nearest vacancy on the local electric field gradient is probably negligible.

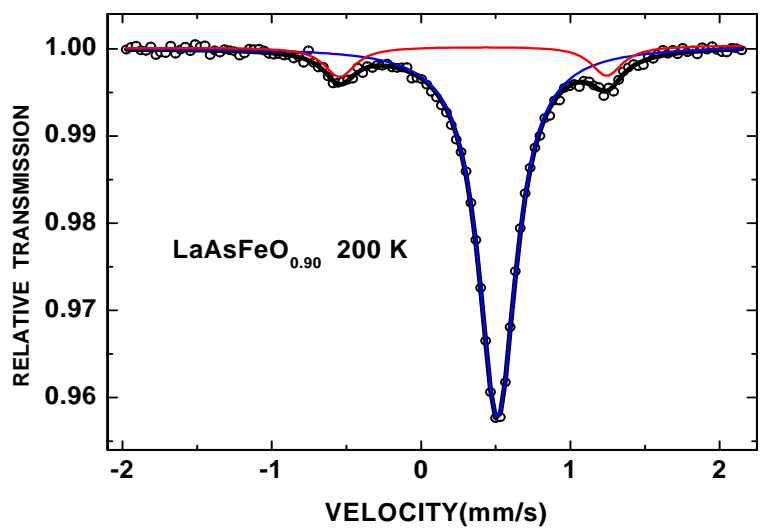

Fig.1. The Mössbauer spectrum of $LaAsFeO_{0.9}$ at 200 K.

b)   The system CeAsFeO$_{0.84}$F$_{0.16}$

CeAsFeO$_{0.84}$F$_{0.16}$ was synthesized by solid state reaction in quartz tube filled with Ar gas at 1250 C. For this composition, the sample is supposed to be SC. However, it appears that this system also exhibits SDW magnetic order at 95 K. At 200 K, Fig. 2, it displays a singlet and a very complicated magnetic spectrum of a foreign phase, which can be identified with high certainty to be Fe$_2$As [2]. The relative abundance of the foreign phase is 17±2%.

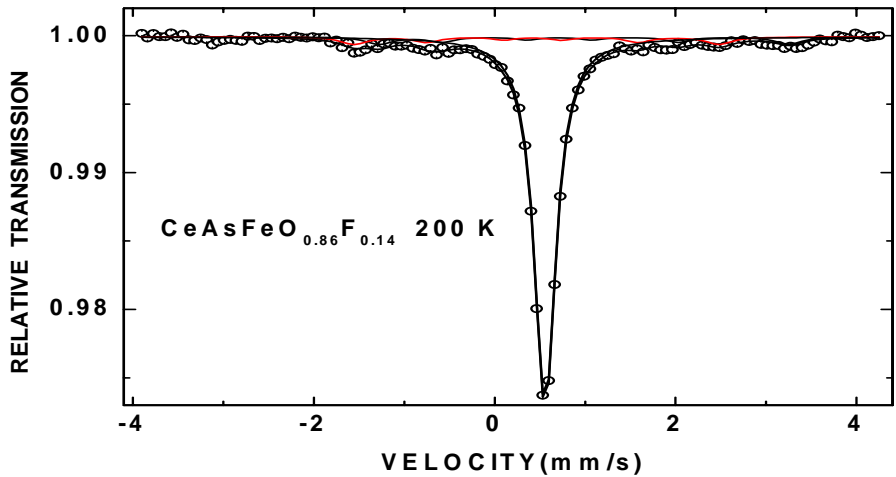

Fig. 2. The Mössbauer spectrum of CeAsFeO$_{0.84}$F$_{0.16}$ at 200 K.

c) The system SmAsFeO

All Sm-Based samples reported here, were prepared under high pressure (6 GPa) at 1300 C [6]. SmAsFeO is SDW system below 140-150 K, however the 200 K Mössbauer spectrum, Fig. 3, shows the presence of a foreign phase of 13±2 % intensity. This impurity quadrupole doublet (I.S.= 0.54(3) mm/s and QS=0.62(2) mm/s) can be identified with full certainty to be FeAs [3].

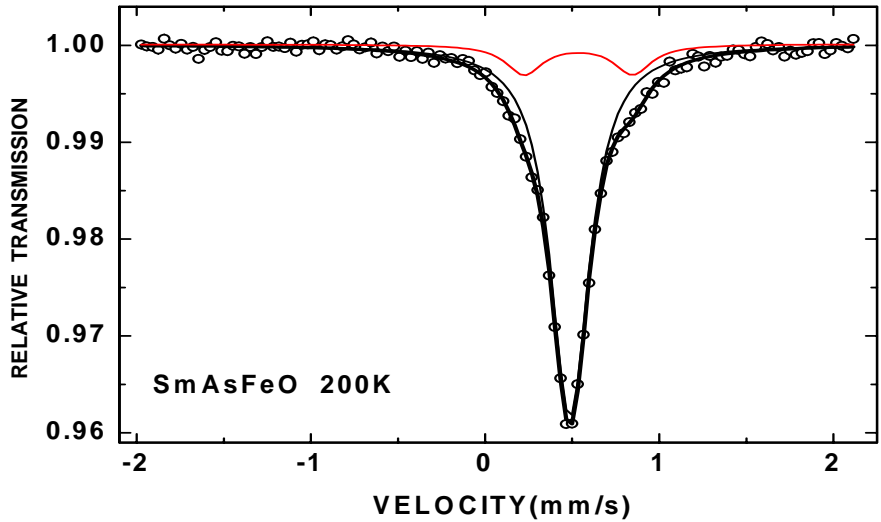

Fig. 3. The Mössbauer spectrum of SmAsFeO at 200 K.

d) The superconducting SmAsFeO$_{0.85}$ and SmAsFeO$_{0.9}$F$_{0.1}$ samples

Both materials are SC at T$_c$=52.6 K and 52 K respectively [6]. However the doublet shown in Figs 4-5 at 95 and 90 K (I.S.= 0.60 mm/s and QS= 0.66 mm/s) which accounts for 15% and 50% respectively, is attributed to FeAs. Both samples exhibit magnetic order at 4.2 K. FeAs orders magnetically at T$_N$=77 K [7], thus the magnetic order in Figs. 4 -5 is presumably due to FeAs as an extra phase. This statement excludes our previous claim for coexistence of SC and magnetic order in this material [8] and also μSR studies in SmAsFeO$_{0.82}$F$_{0.18}$ which observed complex magnetic correlations at low temperatures and relate them to the SC state.[9]

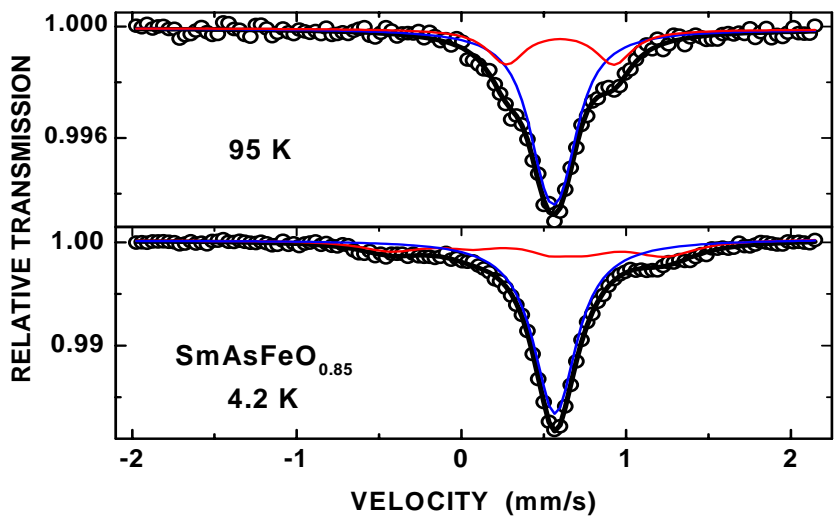

Fig. 4. The Mössbauer spectra of SmAsFeO$_{0.85}$ at 95K and 4.2 K.

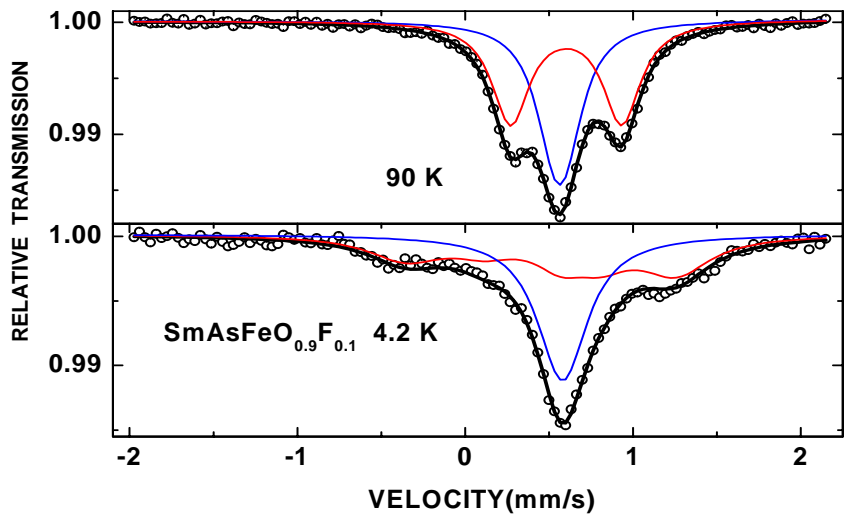

Fig. 5. The Mössbauer spectra of SmAsFeO$_{0.9}$F$_{0.1}$ at 90K and 4.2 K.

e)   The systems Sr$_{0.9}$K$_{0.1}$Fe$_2$As$_2$ and Sr$_{0.6}$K$_{0.4}$Fe$_2$As$_2$

The Sr$_{0.9}$K$_{0.1}$Fe$_2$As$_2$ and Sr$_{0.6}$K$_{0.4}$Fe$_2$As$_2$ samples were synthesized by thoroughly mixing and pressing stoichiometric amounts of the constituents and sintered within a welded Nb container at 900 C [10]. These systems are SDW and superconducting (T$_C$=36 K) materials, respectively. However Sr$_{0.9}$K$_{0.1}$Fe$_2$As$_2$ contains a high concentration (38%) of the magnetic Fe$_2$As (T$_N$=353 K) as an impurity phase (Fig. 6), which is magnetically ordered even at room temperature. At 4.2 K (Fig. 7) the superconducting Sr$_{0.6}$K$_{0.4}$Fe$_2$As$_2$ sample contains 35% of the Fe$_2$As as an impurity.

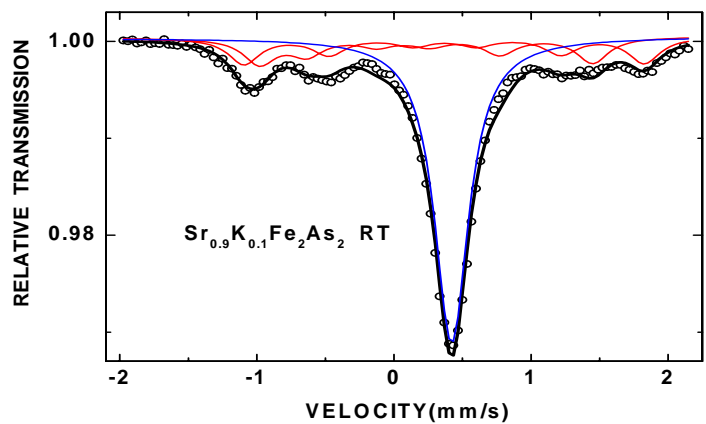

Fig. 6.  The Mössbauer spectrum of Sr$_{0.9}$K$_{0.1}$Fe$_2$As$_2$ at room temperature.

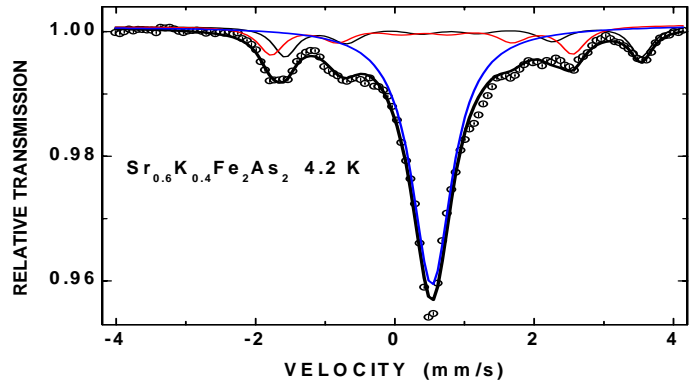

Fig. 7. The Mössbauer spectrum of $Sr_{0.6}K_{0.4}Fe_2As_2$ at 4.2 K.

**Conclusions**

The research reported here proves that one has to be more careful in preparing polycrystalline materials of these superconducting-spin density wave materials, the amounts of impurities in the studied samples reached even 50%. The Fe-As as extra phases are obtained regardless the preparation method of the desired materials. Such materials may lead to wrong conclusions. Our research shows that Mössbauer spectroscopy of $^{57}$Fe in these materials, even at room temperature, can easily discover the presence and amounts of Fe containing foreign phases.

**Acknowledgments**: This research is partially supported by the Israel Science Foundation (ISF, 2004 grant number: 618/04) and by the Klachky Foundation for Superconductivity.